# A Framework for Designing Teleconsultation Systems in Africa

Rowena L. Luk
Intel Research

Melissa Ho
University of California, Berkeley

Paul Aoki
Intel Research

**Abstract:**

*All of the countries within Africa experience a serious shortage of medical professionals, particularly specialists, a problem that is only exacerbated by high emigration of doctors with better prospects overseas. As a result, those that remain in Africa, particularly those practicing in rural regions, experience a shortage of specialists and other colleagues with whom to exchange ideas. Telemedicine and teleconsultation are key areas that attempt to address this problem by leveraging remote expertise for local problems. This paper presents an overview of teleconsultation in the developing world, with a particular focus on how lessons learned apply to Africa. By teleconsultation, we are addressing non-real-time communication between health care professionals for the purposes of providing expertise and informal recommendations, without the real-time, interactive requirements typical of diagnosis and patient care, which is impractical for the vast majority of existing medical practices. From these previous experiences, we draw a set of guidelines and examine their relevance to Ghana in particular. Based on 6 weeks of needs assessment, we identify key variables that guide our framework, and then illustrate how our framework is used to inform the iterative design of a prototype system.*

## INTRODUCTION

Many studies have already shown the value of store-and-forward telemedicine in serving populations limited in financial and medical resources [12, 9]. In particular, telemedicine has the potential to play a central role in health care provision to developing regions, which suffer from conditions such as intermittent electricity, limited bandwidth, and lack of equipment. This can be a particularly effective solution for allowing health care professionals to communicate with each other, since direct patient-provider communication requires much richer information flows to account for the difference in knowledge baseline.

In this paper, we survey the existing literature on store-and-forward telemedicine in developing regions and identify five unique teleconsultation systems that were deployed with practicing doctors, documented, and evaluated. These systems have made significant contributions to the field by verifying that store-and-forward telemedicine is a viable and cost-effective solution; however, they also demonstrate areas of further research which still must be addressed. For instance, one telemedicine system had a turnaround time of over 11 days for a new referral in approximately 10% of 1510 cases [11]. Given the time-critical nature of many of these cases, it is clear that these systems require serious consideration of the 'social engineering' aspects of the system, which determine how and how often users interact with the system. Moreover, most of the systems developed to date have involved isolated deployments between specific individuals or institutions. We propose a framework to clarify some of the implicit social engineering aspects of this these systems, to prevent future systems from repeating the same mistakes but also to clarify challenges in scaling up these systems to larger deployments. This framework is a way of consolidating the success stories and lessons learned in a generalized enough fashion that it can be applied to a variety of future systems. Finally, we examine the relevance of these guidelines to Ghana in particular, based on six weeks of field research, and discuss the iterative design of a prototype system for teleconsultation there.

## METHODS

The following is not a systematic review of store-and-forward telemedicine in the developing world, but rather an attempt to gather a diversity of perspectives on the topic from a variety of rural locales. Over a hundred papers were identified on these topics by searching through Medline and Google Scholar for instances of "store-and-forward telemedicine", "teleradiology," "telepathology," "e-mail telemedicine," and "teleconsultation," to name a few. Moreover, we solicited references from a variety of experts in the field, many of whom were responsible for building and deploying their own teleconsultation systems. Papers were selected for further study if they fulfilled the following criteria:

1) The papers discuss the design of a system for health care professionals to exchange information about particular cases.
2) The application is targeted towards developing regions or regions with comparable local

**Table 1 – Summary of Store-and-Forward Telemedicine Deployments in Developing Countries**

| Name (Specialization) | Location | Deployment | Time Period | Monthly Usage | Turnaround Time | Comments |
|---|---|---|---|---|---|---|
| iPath (Pathology) | Solomon Islands | National Referral Hospital (NRH), serving 450 000 people across 900 islands. 8 pathologists on call. | 24 mo. | 14 | 8.5 days | Currently Active. ~275 cases submitted per month |
| Labrador Telemedicine Project (General) | Rural Canada | Black Tickle, serving 210 people 1 resident nurse 2 consulting physicians on call. | 12 mo. | 7 | n/a (urgent cases were phoned in) | Use of a commercial medical case creation software |
| EHAS (General) | Peru | 1 provincial hospital 7 health centers 31 health posts | 9 mo. | 20 /facility | n/a (email system frequently down) | Strong use for referral (205 emergency transfers in 9 months) |
| Swinfen Charitable Trust (Various) | Bangladesh | Centre for the Rehabilitation of the Paralysed in Bangladesh, population: 120 million. 5 specialists on call. | 12 mo. | 2 | Approximately 1 day | Today, 59 hospitals in 22 countries and 144 specialists are using this system (Wootton 2005). |
| n/a (General) | Cambodia | 3 hospitals in Rovieng, Cambodia | 28 mo. | 9 | 1 day | A nurse visited a local health center and transmitted observations to doctors, for feedback and treatment the subsequent day. |

conditions in terms of limited human and financial resources.
3) Communication is non-real-time.
4) The application was used by real doctors and evaluated with their feedback.

Using the above criteria, ten studies were identified that involved the unique design of teleconsultation systems in the developing world (or rural parts of the Western world). Table 1 summarizes the studies in chronological order [12, 2]. Note that, given the emerging nature of this field of research, many of the studies are only at preliminary stages. Because the systems are designed differently, the studies use different methods, and the target population varies tremendously in terms of a variety of local factors, it would not be sensible to try to combine all the results. However, the table helps us to summarize what we do and do not know from these studies. We subsequently elaborate on some of the key lessons learned from a few of them.

Drawing primarily from the studies summarized above, but also from six semi-structured interviews with researchers in different telemedicine projects, we propose a framework for the design of teleconsultation system. We then apply this framework to the iterative design process of a teleconsultation system for physicians in Ghana. Our design process is informed by six weeks of fieldwork in nine medical institutions to assess the local conditions, wherein we employ semi-structured interviews, participant observation, and focus groups with a variety of physicians and associated staff.

**FRAMEWORK**

From this existing literature, we propose a social framework to guide the design of future systems, which we base upon three ground principles: *accountability*, *connectedness*, and *respect*.

These are features that the existing system designs fail to address directly, although most of them make provisions for each element in one way or another. Accountability refers to the fact that designers need to identify specific incentives for doctors to provide quality, continuing input into the system, while those asking for consultation need a reason to provide follow-up service. Connectedness refers to the social elements that engage doctors to adopt the system in the first place and to turn to it when they are seeking new information. Respect refers to the relationship between doctors on this system; in telemedicine, where participants are often located on the other side of the world from each other, one must explicitly take into consideration broad cultural and resource differences when facilitating communication between health care providers.

Following are a few key areas of design which draw most strongly from the above principles.

## APPLICATION OF THE FRAMEWORK

While there are a variety of design decisions that go into these systems, we apply our framework to four areas that contributed significantly to the success of the project but areas in which we also witnessed significant variability. Those four areas are:

1) Choosing an appropriate locale to support these systems
2) How to arbitrate case assignment between health care providers
3) Tackling the differential access to information and physical resources between providers
4) What kind of follow-up or evaluation methods to use

### Choosing an Appropriate Locale

Because this type of store-and-forward remote consultation is only one of a broad range of telemedicine tools, it is important to identify key traits of an area that lend itself towards this particular practice. Clearly, this is a good option if there are no others available [1]. However, systems like this flourish particularly in populations that suffer severe professional isolation and could benefit not only from the professional but also the personal benefits of information technology. Existing connectedness between providers can be a key criterion for adoption of an unfamiliar technology, and, moreover, allows users to openly ask small, non-case-specific questions which are still important to the overall operation of a rural clinic [7]. Rural providers often think of these links as a "one-stop shop" for medical information they cannot easily access, such as "the best method of pain relief in pediatric burns, in a hospital with very few drugs" or "what brand, length and gauge of needle would be suitable for stereotactic core biopsy," according to a personal correspondence with Richard Wootton of the Centre for Online Health, Queensland, Australia. This social element for adoption can be seen in the EHAS project in Peru, one of the most active teleconsultation projects we studies, which recorded an average of 19 personal communications per month, a significant increase from previous personal communication at a rate of twice a month.

### Case Assignment

If there is one critical aspect of this system that is most contingent on the social engineering of the system, it is case assignment. If and how cases are assigned from a provider seeking advice to one giving it significantly impacts the frequency, relevance, and applicability of information acquired. Accountability is really the key element of this aspect of the design, for information-providers need an incentive to respond to questions quickly.

iPath presents an important innovation in this field when it introduced the idea of the "Virtual Institute of Pathology" in the Solomon Islands [2. Previously, iPath provided an open, bulletin-board style system whereby specialists willing to provide services would select and answer cases of interest to them. With the Virtual Institute, specialists around the world were organized into collectives, which were monitored by one individual "on call" at any time who would moderate the list and ensure that all requests for information were answered in a timely fashion. This institute also enabled "local" discussion among the experts before a final recommendation was transmitted back to the asker. By this system, one and only one individual was accountable for seeing a consultation through, and because of this, the turnaround time for a first response dropped by about 70% (from 28 hours to 8.5 hours). While small-scale collaborations between one or two people on a variety of cases can provide remarkable turnaround times in less than 24 hours, the demand clearly exists for these systems to scale to larger deployments, and when that happens there needs to be a strong sense of accountability in the system [1, 13]. Of course, the tradeoff with an assignment system is that there needs to be some way of identifying the individual most qualified to provide the information, while at the same time specialists are the most qualified to determine the extent of their own specialization. Thus, there will clearly be times when the value of the information is such that it is worthwhile to wait for the appropriate specialist to step up to the plate. However, as is the case in physical hospitals, where there is always one doctor "on call" to deal with whatever cases appear in the emergency ward, so in the interest of time must there be one consultant responsible for providing a timely follow-up. If nothing else, a doctor in a given field of work should have a clear idea of what he does not know. This accountability is even more important as one considers other potential applications for store-and-forward telemedicine, such as electronic medical health records or the building of repositories of local case information; without follow-up data on the outcome of consultations, these systems lose most of their value.

### Differential Access to Resources

Because we are targeting telemedicine within developing regions, it is often the case that information or expertise is sought from another country where resources and medical education are more common. As such, most systems must address issues of cultural conflict, differences in education, access to resources, and local medical conditions, which can result in a consultant requesting unrealistic follow-up diagnostics or limited pertinence of any recommendations [6, 1]. Some projects addressed this problem by cutting specialists out of the loop entirely, and referring cases almost entirely to primary care physicians who are less likely to request expensive follow-up procedures [10]. However, there is no doubt that specialist knowledge needs to be maintained in the system, since their knowledge is a key value addition.

In addressing this problem, we emphasize that the most important element of this collaboration is respect for the differences in environment and the provision of alternative solutions. Training is an essential element, not only to make sure users are comfortable with the system but also to ensure their understanding of the local context. Strategies to address the differences include consultants providing a variety of diverse treatment options, so that the local health care professional can make a judgment call based on what is locally available [12]. Whatever the local conditions, however, foreign consultants must respect that the local provider is an expert on local conditions, making use of what resources he or she has access to, and the consultant must be willing to frame his answers within local constraints.

**Evaluation**

Information-seekers need incentives to follow-up on cases with the outcome of particular recommendations. This is important for a variety of reasons including

1) to build strong medical health records
2) to develop repositories of local case information [3]
3) to provide good feedback and a learning opportunity to the information-providers [9, 1].

Accountability, in this case, applies to the information-seekers; that they know and can see that they are asking for advice from other busy professionals and that they themselves are responsible for following-up on those advices. Without this follow-up, a large fraction of the cases studied were not able to provide comprehensive system evaluations to improve future work [1, 12]. Strong connectedness is one way of addressing this problem, for if the providers have other reasons to dialogue then invariably the results of previous consultations will come up in subsequent cases. However, no compelling system exists to date that provides accountability among the information-seekers.

## APPLICATION TO GHANA

In the developing regions of the world, brain drain is an especially significant problem in Ghana, with 30 out of every 100 Ghanaian doctors practicing in the United States, the United Kingdom, Canada, or Australia [5]. Given the strong social connections within the small medical elite working in Ghana, the professional isolation of rural doctors, and poor connectivity in regional hospitals, we anticipate doctors working in regional hospitals or rural heath centers would have significant interest in a potential teleconsultation project.

Drawing on fieldwork we did in Ghana in 2005-2006, we have previously identified four key themes in the Ghanaian health care system [8]. These are:

1) Failures in the existing patient referral system
2) Healthcare practices that largely do not involve physical contact
3) Perception of telemedicine as valuable but expensive
4) A shortage of medical specialists.

Motivated by these conditions, we draw on the principles of accountability, connectedness, and respect within our framework to flesh out many of the design decisions in our system.

## PROTOTYPE SYSTEM

Remote Asynchronous Communication for Healthcare (REaCH) is designed to address the failures in the patient referral infrastructure, by creating an online system whereby rural doctors in Ghana can enter case information, in the form of text or images, and send, or "assign" that case to a specialist in one of the central hospitals. The lack of physical contact in existing practices is a good indicator of the potential value of our system, while the store-and-forward element of our system makes it a cheap, low-cost tool that does not require high-speed internet connections and works over e-mail as well as through a web-based portal. Moreover, we take this one step further and leverage the DTN (delay-tolerant networking) architecture, which would even allow computers not connected to an intranet or internet through traditional media to be updated via a USB key [4]. We address the shortage of medical specialists in the country by opening up this system to volunteer specialists around the world, who would provide a second-tier group of consultants if those in Accra do not feel they have the time or expertise to handle a particular case.

The social framework we outlined earlier plays a key role in solidifying many of our design decisions. Because Ghana only has two medical schools, both with graduating classes of less than a hundred students, the medical community is highly interwoven, both professionally and socially. This connectedness is a key element which will drive this system forward, and which we hope to leverage even with foreign consultants, by drawing, where possible, from the Ghanaian diaspora overseas. In fact, we anticipate leveraging this social aspect even within the user experience design, by allowing doctors to easily link and communicate with informal colleagues in a lightweight, non-case-specific manner.

To create accountability for specialists providing expertise in this system, we allow information-seekers to assign their cases to a few potential candidates, while the system does the final assignment of the case to a single candidate (in order to balance the case load automatically among specialists). Such a system can potentially leverage whatever social connections already exist between information-seeker and information-provider, and it will always ensure that the seeker knows who the provider is and can associate responsiveness to the individual instead of the system overall.

On the flipside, information-seekers are accountable in the system through the fact that each will have a publicly available profile with uniquely identifying data. Thus, a specialist wishing to know the outcome of a particular case knows exactly who to contact and how.

Respecting the existing health practices in Ghana, we choose not to circumvent the existing referral infrastructure and allow all cases to pass through a first assignment tier with local Ghanaian specialists. It is only in the case of work overload that these cases are passed onwards to foreign specialists. Where possible, all participants in this system will have field knowledge of medical practice in Ghana, so that they are intimately familiar with the people and conditions in which they are working. In cases where this is not possible, a full-day training session will be developed in conjunction with local doctors and provided to specialists abroad.

## CONCLUSION

We endeavor in this paper to describe a framework for designing teleconsultation systems, suggesting that such projects should embody accountability, connectedness, and respect. This framework draws from a literature survey of existing telemedicine applications of the same genre. Based on this, we design and deploy a teleconsultation system enabling better informed referrals between rural doctors and specialists in Ghana.

## ACKNOWLEDGMENTS

This work was made possible through the contributions of Holly Ladd, Lady Pat Swinfen, Richard Wootton, Victor Patterson, Dr. Lionel Dibden, Linda Ogilvie, and the many doctors who have so graciously volunteered their time to give us insights into telemedicine and the design process.

**Address for correspondence**

Rowena L. Luk
Intel Research Berkeley
2150 Shattuck Avenue, Penthouse Suite
Berkeley, CA 94704
United States
rluk@sims.berkeley.edu